\documentstyle[psfig,aaspp4,12pt]{article}

\lefthead{J.~M.~Miller et al.}  \righthead{Relativistic Iron Emission
and Disk Reflection in Galactic Microquasar XTE~J1748-288}

\received{Date}
\revised{Date}
\accepted{Date}

\journalid{vol}{date}
\articleid{1}{4}
\paperid{id}

\cpright{AAS}{1999}
\ccc{x}

\newcommand{\xtej}{\mbox{XTE~J1748$-$288}}

\newcommand{\etal}{\mbox{et al.}}

\newcommand{\ascaz}{\textit{ASCA\/}}
\newcommand{\rxte}{\textit{RXTE}}
\newcommand{\rxtez}{\textit{RXTE\/}}

\newcommand{\ergcms}{ergs cm$^{-2}$ s$^{-1}$}

\newcommand{\Msun}{\mbox{$M_{\sun}$}}
\newcommand{\rah}{\mbox{$^{\rm h}$}}
\newcommand{\ram}{\mbox{$^{\rm m}$}}

\begin{document}

\title{Relativistic Iron Emission and Disk Reflection in Galactic
Microquasar \xtej}

\author{J.~M.~Miller,\altaffilmark{1}
        D.~W.~Fox,\altaffilmark{1}
        T.~DiMatteo,\altaffilmark{2,5}
	R.~Wijnands,\altaffilmark{1,5}
        T.~Belloni,\altaffilmark{3}
	D.~Pooley,\altaffilmark{1}
	C.~Kouveliotou,\altaffilmark{4}
	W.~H.~G.~Lewin\altaffilmark{1}
	}

\altaffiltext{1}{Center~for~Space~Research and Department~of~Physics, Massachusetts~Institute~of~Technology, Cambridge, MA
        02139--4307; jmm@space.mit.edu, derekfox@space.mit.edu, rudy@space.mit.edu, davep@space.mit.edu, lewin@space.mit.edu}  
\altaffiltext{2}{Harvard--Smithsonian Center for Astrophysics,
        Harvard University, Cambridge MA 02138; tdimatte@cfa.harvard.edu}
\altaffiltext{3}{OAB Brera, Italy;  belloni@merate.mi.astro.it}
\altaffiltext{4}{NASA MSFC, SD-50, Huntsville, AL 35812; kouveliotou@eagles.msfc.nasa.gov}
\altaffiltext{5}{\it Chandra fellow}


\authoremail{jmm@space.mit.edu}

\label{firstpage}

\begin{abstract}
We report evidence for an Fe~K$_{\alpha}$ fluorescence line feature
and disk reflection in the Very High, High, and Low State X-ray
spectra of the galactic microquasar \xtej\ during its June~1998
outburst.  Spectral analyses are made on data gathered throughout the
outburst by the {\em Rossi X-ray Timing Explorer\/} Proportional
Counter Array.  Gaussian line, relativistic disk emission line, and
ionized disk reflection models are fit to the data.  In the Very High
State the line profile appears strongly redshifted, consistent with
disk emission from the innermost stable orbits around a maximally
rotating Kerr black hole.  In the High State the line profile is less
redshifted and increasingly prominent.  The Low State line profile is
very strong ($\sim$0.5~keV equivalent width) and centered at
6.7$\pm$0.10~keV; disk line emission model fits indicate that the
inner edge of the disk fluctuates between $\sim$20 and
$\sim$100~$R_{\rm g}$ in this state.  The disk reflection fraction is
traced through the outburst; reflection from an ionized disk is
preferred in the VHS and HS, and reflection from a relatively neutral
disk is preferred in the LS.  We discuss the implications of our
findings for the binary system dynamics and accretion flow geometry in
\xtej.
\end{abstract}

\clearpage

\section{Introduction}

Quasars and other Active Galactic Nuclei (AGN) are generally thought
to be powered by accretion onto supermassive black holes at the
centers of galaxies.  They are well-studied across the full
astronomical bandpass, from radio to gamma-ray wavelengths.  At X-ray
energies, attention in recent years has focused on the broad
Fe~K$_{\alpha}$ fluorescence line seen in many AGN of the Seyfert~1
type (for a recent review, see Fabian \etal\ 2000).  At the high
spectral resolutions first achieved with the \ascaz\ satellite, the
line in these sources is revealed to have a strongly asymmetric
profile that is consistent with emission from the innermost radii of a
relativistic accretion disk (Tanaka \etal\ 1995; Mushotzky \etal\
1995; Nandra \etal\ 1997).  The theory for the emission mechanism
holds that the X-rays originate by inverse-Compton processes acting in
a hot corona above a cold (weakly ionized) accretion disk that
produces thermal emission in the optical and ultraviolet; thus the
theory is bolstered by observations of a reflected continuum above
$\sim$10~keV that flattens the intrinsic inverse-Compton power-law.

Within our Galaxy, radio and X-ray observations over the course of the
past six years have identified a population of X-ray binaries, the
galactic microquasars, that are similar to AGN in a number of
important respects.  First, these sources exhibit relativistic
radio-emitting outflows (e.g., Mirabel \& Rodriguez 1994) analogous to
the megaparsec-scale radio jets of AGN.  Second, the sources are
thought to harbor black hole primaries: they exhibit the canonical
X-ray spectral states of black hole candidates (BHCs; see discussion
below), and when dynamical mass estimates have been obtained they have
supported this hypothesis (e.g., Orosz \& Bailyn 1997).  Finally, the
sources exhibit strong X-ray variability on a broad range of
timescales (hundreds of hertz to hundreds of days), also analogous to
AGN.  The name ``microquasar'' refers approximately to the relative
mass difference between the two classes of object: the $\sim$10~\Msun\
masses of the microquasars, on the one hand, and the
10$^7$--10$^9$~\Msun\ masses of AGN, on the other.  This relative mass
scale in turn is expected to set the relative variability timescale.

In addition to the properties mentioned above, an important common
characteristic of the microquasars (without yet an analogue in AGN)
are X-ray quasi-periodic oscillations (QPOs) in the 30--300~Hz
frequency range.  These QPOs have been identified in the sources
GRS~1915$+$105 (Morgan, Remillard, \& Greiner 1997), GRO~J1655$-$40
(Remillard 1997), XTE~J1550$-$564 (Remillard 1999), \xtej\ (Fox \&
Lewin 1998; Revnivtsev, Trudolyubov, \& Borozdin 1999), and
4U~1630$-$47 (Cui \etal\ 1999a).  The phenomenology of these QPOs is
remarkably rich and continues to challenge both the observers seeking
to characterize them (e.g., Reig \etal\ 2000; Sobczak \etal\ 2000),
and the theorists seeking to explain them (e.g., DiMatteo \& Psaltis
1999; Lehr, Wagoner, \& Wilms 2000).

Not surprisingly, the microquasars have also been found to be a rich
arena in which to study accretion physics and geometry via X-ray
spectral features.  \textit{Rossi X-ray Timing Explorer\/} (\rxte;
Bradt, Rothschild, \& Swank 1993) Proportional Counter Array (PCA;
Jahoda \etal\ 1996) observations of XTE~J1550$-$564 are well-fit by
including a Gaussian line feature at 6.5 keV with fixed width of 1.2
keV (FWHM) (Sobczak \etal\ 1999).  Two line features, one at $\sim$5.7
keV and one at $\sim$7.7 keV, were observed in the 1996 outburst of
4U~1630$-$47 (Cui, Chen, \& Zhang, 1999).  The two lines are
attributed to Doppler shifting of neutral Fe~K$_{\alpha}$ in either a
Keplerian accretion disk or bipolar outflow.  Balucinska-Church \&
Church (2000) report likely Fe~K$_{\alpha}$ emission from four spectra
taken from GRO~J1655$-$40.

\xtej\ was discovered with the \rxtez\ All Sky Monitor (ASM; Levine
\etal\ 1996) on 4~June~1998 (Smith \etal\ 1998).  Radio observations
undertaken with the Very Large Array (VLA) revealed an unresolved
radio source with position coincident to that of the X-ray source
(R.A. $17\rah 48\ram 05\fs 06$, Dec.$\ -28^{\circ} 28\arcmin 25\farcs
8$, J2000; Hjellming \etal\ 1998a).  Subsequent radio observations
revealed an extended source, exhibiting components with proper motion
of 20--40 mas~day$^{-1}$ (Rupen, Hjellming, \& Mioduszewski 1998).
Based on a 21-cm HI absorption measurement the distance to the source
was estimated to be $\geq$8~kpc, implying that the intrinsic velocity
of the jet components must be higher than 0.93$c$ (Hjellming \etal\
1999b).  QPOs with centroid frequencies of $\sim$0.5 Hz and $\sim$32
Hz were reported in the power density spectrum (PDS) from 6~June~1998
(Fox and Lewin, 1998).  The combination of relativistic radio jets and
X-ray QPOs secured the identification of \xtej\ as a Galactic
microquasar.  Multiple observations of the source were made with
\rxtez\ (Revnivtsev, Trudolyubov, \& Borozdin 1999) and \ascaz\
(Kotani \etal\ 2000); the source flux decayed below the ASM detection
limit in September 1998.

The evolution of black hole X-ray binary (BHXB) outbursts is
generally described in terms of several canonical spectral states,
distinguished by the relative contribution from low energy and high
energy X-rays within the 1--10~keV band. Typically, low energy X-rays
(1--2~keV; the soft component) dominate the emission of a BHXB
outburst near maximum, and as the outburst evolves the soft component
grows weaker.  Higher energy X-rays (5--10 keV; the hard component)
become stronger and dominate as the overall flux decreases in the late
stages of a BHXB outburst.

The canonical BHXB spectral states are the Very High State (VHS), the
High or Soft State (HS), the Intermediate State (IS), the Low or Hard
State (LS), and the Quiescent State (QS).  In the LS, the X-ray
spectrum is nearly a pure power-law, with photon index 1.5--2.5, often
extending to hundreds of keV (Tanaka \& Lewin 1995).  In the HS, the
1--10~keV flux is an order of magnitude higher than in the LS due to
the addition of a strong soft component.  In the VHS the soft
component is stronger still and may be accompanied by a substantial
hard component; in addition, QPOs are often observed in this state.
The IS is a transitional state between the LS and the HS (Mendez \&
van der Klis, 1997), not observed in all sources.  Finally, the QS
state is typified again by a non-thermal spectrum, with a flux level
several orders of magnitude below the lowest (LS) outburst flux.

Although the temporal and spectral behavior of BHXBs has been widely
studied, the mechanism driving the state transitions is still unknown.
The quasi-thermal radiation present in the VHS, HS, and IS is
typically modeled as thermal emission from an optically-thick,
geometrically-thin, multi-temperature accretion disk (Mitsuda \etal\
1984).  The power-law component, meanwhile, is generally attributed to
inverse-Compton processing of lower-energy disk photons in an
optically-thin corona above the accretion disk (Sunyaev \& Titarchuk,
1980).  If signs of hardening in the spectrum above $\sim$10~keV are
present and accompanied by a fluorescent Fe~K$_{\alpha}$ line, then --
much as in AGN -- these can be interpreted as the signature of Compton
reflection from the cold matter in the disk (George \& Fabian 1991).

Reflection features in the LS of BHXBs, however, are generally much
less prominent than in AGN (Zycki \etal\ 1999), so it has been argued
that at least in the LS, the geometrically-thin disk does not extend
inward to the innermost stable orbit, but is truncated at hundreds or
thousands of Schwarzschild radii (Esin \etal\ 1997).  In the HS, the
optically thick disk is postulated to move inwards so that the
majority of the dissipated energy emerges in the form of a
blackbody-like spectrum.  Measuring reflection spectra in BHXBs may
therefore be helpful in understanding the physical properties of BHXB
accretion geometries.
  
Previous analysis of \rxtez\ observations of \xtej\ revealed
Fe~K$_{\alpha}$ emission in the LS spectra of this source (Revnivtsev,
Trudolyubov, \& Borozdin 2000).  Our fits to the same data revealed a
strong LS iron feature, and in addition we found that it was possible
to trace the feature backwards into the HS, and even the VHS.  Here we
present the results of 22 X-ray spectral observations spanning
$\sim$120 days of the 1998 outburst of \xtej.  We investigate
relativistically-skewed emission features and disk reflection in the
spectra from this source.

\section{Observations and Spectral Analysis}
\subsection{Observations}
We include all of the publicly-available \rxtez\ TOO data between peak
bolometric luminosity and quiescence in our analysis.  This embodies
22 pointed observations, effectively sampling the different spectral
states across the outburst; see Table~1 for times and ObsIDs of these
observations.  The standard FTOOLS v.4.2 package was used to reduce
the Standard-2 PCA data from these observations.  All observations were
dead-time corrected, and the background was calculated using the
Very Large Events (or Bright Source) option within PCABACKEST.

\placetable{1}

\subsection{Gain Drift and Response Matrices}
The \rxtez\ PCA instrument (Jahoda \etal\ 1996) consists of 5
individual proportional-counter detectors, the proportional counter
units (PCUs).  Due to gas exchange between propane and xenon layers in
the detectors, the gain of the individual PCUs drifts.  PCUs 0, 1, and
4 have very similar gain drift patterns, and a better overall
performance than PCUs 2 and 3 (Sobczak \etal\ 1999; Balucinska-Church
\& Church 2000).  We have chosen to fit data from all layers (all
propane layers and all xenon layers) of PCUs 0, 1, and 4
simultaneously.

Analysis of the data from an May~1998 observation of the Crab nebula
with the response matrices calculated via the Ftool PCARMF (v3.5)
shows substantial residual trends.  These deviations are thought to
represent defects in the response matrices themselves as the Crab is
known to have a featureless power-law spectrum.  We use response
matrices generated by PCA calibration specialist Keith Jahoda that
eliminate these residuals.  They are available through the Goddard
Space Flight Center ftp archive at
\verb+ftp://lheaftp.gsfc.nasa.gov/pub/keith/response_matrices_v2.2.1_v80.tar+.
A comparison of the data/model ratio using PCARSP-generated and
Jahoda-generated (hereafter ``static'') response matrices, presented
in Fig.~1, reveals that the static response matrices yield a much
cleaner fit to the Crab data in the 3.5--10.0 keV band.  For this
reason, we use the static response matrices for all spectral fits.

\placefigure{1}

\subsection{Calibration via the Crab Nebula}
We have analyzed every non-slewing observation of the Crab nebula from
May to October~1998 to check for variations in the performance of the
static response matrices over time.  

These Crab fits were used to select an appropriate energy range for
precision analysis of the data from \xtej.  The bins between 2.0 and
3.5~keV varied drastically and randomly in fits to these Crab
observations: the $\chi^{2}$ statistic for these bins consistently
fell in the range of $\sim$40--100.  Therefore, we set a lower energy
limit of 3.5~keV for our analysis.  Above $\sim$20~keV, neither type
of response matrix satisfactorily fits the Crab.  As this energy is
high enough to constrain reflection features fit to the data, we fix
20~keV as the upper limit of our fitting range.

The addition of 1\% systematic errors has become a standard practice
when reducing PCA spectral data (see, e.g., Cui, Chen, \& Zhang, 1999;
Sobczak \etal\ 1999).  Fits to the set of Crab observations mentioned
above with this 1\% systematic error produced artificially low
$\chi^{2}$ values.  We find that a systematic error of 0.75\% yields
fits to the Crab observations with $0.99<\chi^{2}/(dof)<1.0$; we
therefore adopt this systematic error value for our fits.  This step
adds validity to the measures of statistical merit based on
$\Delta(\chi^{2})$ that we employ in our analysis.

\subsection{PCA Calibration Sources}
Onboard the PCA are several pellets of radioactive Americium,
$^{241}$Am, that are used for in-flight calibration.  Data from these
sources is accumulated every 128 seconds and telemetered as part of
the Standard1 data.  As the line emission energies of the source are
precisely known, any drift in the measured energy of the lines over
the lifetime of the satellite can be directly attributed to changes in
the detector response.

We collected all of the calibration source data from the first
(4~June) and last (26~September) days on which \xtej\ was observed;
analyzing this data in channel space we find that the centroids and
widths of the $^{241}$Am lines agree to $<$1\% and $<$2\%,
respectively, for these two days.  We therefore expect that the PCA
energy scale drift is less than 1\% over the time \rxtez\ conducted
these observations.  Assuming a linear scale, this represents a
systematic drift of $\leq$0.06~keV for a line feature centered at
6.0~keV.

\subsection{ASCA-measured $N_{H}$}
As we determined, \rxte's effective sensitivity range does not extend
below 3.5~keV during the time of our observations.  We therefore rely
on \ascaz\ observations to measure the column density reliably.
Kotani \etal\ (2000) report $N_{H} = 9.0 \times 10^{22} cm^{-2}$ for a
power-law index of $\alpha_{pl} = 2.9$, and $N_{H}$ at $6.0 \times 10^{22}
cm^{-2}$ for $\alpha_{pl} = 2.7$, each with an error of approximately
$\pm$30\%.  Both of these ASCA-measured power-law indices agree with
the power-law indices we measure using \rxtez\ data.  We therefore fix
the value of $N_{H} = 7.5 \times 10^{22} cm^{-2}$ for all of our fits.

\subsection{Fitting Method}
Our spectral fits progressed through three increasingly complex
spectral models, a process which we illustrate for two representative
observations (obs.\ 3 and 19, from the VHS and LS, respectively) in
Figure~2.

\placefigure{2}

First, we fit the data with the canonical multicolor blackbody disk
plus power-law model (top and second panels in Figs.~2a and 2b).  The
most striking feature in the data/model ratio plots for this model
(second panel) is a broad excess between 4 and 7~keV for Obs.~3, and a
distinct 6.7~keV line for Obs.~19.  Our second model therefore adds
one (Fig.~2b) or two (Fig.~2a) Gaussian components to the underlying
continuum (third panel).  In the case of Obs.~19 (and other LS
observations), the fit is further improved by addition of a Gaussian
at 8~keV (fourth panel in Fig.~2b); we discuss this feature in more
detail in Sect.~\ref{res:line} below.

Two spectral features are expected to result from the reflection of
hot X-rays from an accretion disk: fluorescent line emission, and a
broad Compton-upscattering excess often referred to as a ``reflection
bump.''  To be self-consistent, our final model (bottom panels in
Figs.~2a and 2b) therefore consists of the multicolor blackbody disk
(Diskbb) component, two Gaussians, and a Pexriv component (Magdziarz
\& Zdziarski 1995).  Pexriv incorporates an incident power-law source
of photons, and the reflection of these photons off an ionized disk.
For all fits with Pexriv, we fix the power-law cutoff energy at
200~keV, and the disk inclination at 45 degrees (as we do for disk
line profiles).  We report measurements of the disk reflection
fraction ($f~=~\Omega/2\pi$) and ionization ($\xi~=~L_{x}/nr^{2}$).
Pexriv reduces to a simple power-law in the absence of reflection.

We report the results of our continuum and Gaussian-line fits in
Tables~2 and 3, respectively.  

\placetable{2}
\placetable{3}

Models for emission line propagation from a relativistic spacetime
generally predict a double-peaked line profile (see, e.g., Dabrowski
\etal\ 1997, Martocchia \etal\ 2000).  Noting that the double-peaked
nature of the soft emission excess in the VHS might be due to
relativistic skewing, we attempted to fit these spectra with a
relativistic line feature, the Laor model from XSPEC (Laor, 1991).
This is a general model and may be applied to Kerr geometries but
reduces to a Schwarzschild geometry in the minimal angular momentum
limit.  In the absence of optical or radio data to constrain the disk
inclination angle, we fix this parameter at 45 degrees.  In addition,
the disk emissivity profile is fixed at $r^{-2}$, the inner emission
radius of the disk to $1.235~R_{\rm g}$ (as for a nearly-maximally
rotating Kerr black hole), and the outer emission radius to
$50.0~R_{\rm g}$.  The gravitational radius $R_{\rm g}$ is $GMc^{-2}$, where
$G$ is Newton's gravitational constant, $M$ is the mass of the compact
object, and $c$ is the speed of light ($R_{\rm g}$ is half of the
Schwarzschild radius).  Measurements of the inner disk radius in the
VHS are reported in Table~4.

\placetable{4}
        
In the LS, models for accretion flow around black holes (Esin \etal\
1997) suggest that the inner edge of the accretion disk does not
extend as far inwards as in the VHS.  Measurements of the inner disk
radius derived from the normalization of the multicolor blackbody disk
component suggest the same phenomenology.  It follows that emission
from the inner disk regions in the LS would not suffer the same degree
of relativistic skewing.  Thus, an additional fit is made to LS
spectra with a model consisting of the ``Diskline'' line emission
model within XSPEC (Fabian \etal\ 1989) and Pexriv.  For these fits,
we fix the disk emission profile to $r^{-2}$ and the disk inclination
to 45 degrees; and the outer disk radius to $10^{4}~R_{\rm g}$.  We fit
for inner disk radius, line centroid energy, and the component
normalization.  The results of fits using this model are reported in
Table~5.

\placetable{5}

\section{Results}

\subsection{Continuum Components}
\label{res:con}
The total 3.5--20.0 keV flux measured using our models peaks at
11.4$\times 10^{-9}$~\ergcms~between outburst days 2 and 8, and is
dominated by the hard power-law component.  The soft multicolor disk
blackbody flux is slightly less than half of the hard power-law flux
in observations 3 and 4, but a smaller fraction in observations 5--8
(see Table~2).  The normalization of the blackbody disk component is
tightly constrained in this state, and is relatively constant.  These
relative flux measurements and very high overall flux level signify
that \xtej\ is in the VHS during observations 3--8.  The disk
blackbody temperature is well-measured, and decreases very slightly
through this state, from 1.57 to 1.26~keV.  The normalization of the
hard power-law component is well-constrained; the spectral index
hardens through the VHS (see Table~2, Figure~4a).

Between observations 8 and 9, the hard power-law flux falls and the
3.5--20.0 keV spectrum becomes dominated by the soft component.  The
multicolor blackbody disk flux increases by a factor of $\sim$4
between these observations.  We interpret this change in continuum
component flux as the VHS to HS transition.  The blackbody disk
temperature decreases from 1.29 to 0.91 keV through this state (see
Table 2).  The normalization of the soft component peaks mid-way
through the HS, dipping most dramatically in observation 15, just
prior to the HS--LS transition.  The power-law spectral index is
softer than in the VHS at the beginning of the HS, but the index
steadily hardens through the HS.  Whereas the soft blackbody disk
normalization peaks mid-way through this state, the hard power-law
component normalization falls steadily throughout (see Table 2).

The multicolor blackbody disk temperature falls by half between
observations 15 and 16, and the soft flux is nearly extinguished.
This marks the HS to LS transition.  Ginga observations of
GS~1124$-$68 (Ebisawa \etal\ 1994) and GS~2023$+$338 (Zycki \etal\
1999) yield LS disk temperatures of $\sim$0.3~keV.  We measure a disk
component with similarly low temperature in observations 16, 17, and
18, but find no evidence of a disk flux in subsequent observations.
It is therefore possible that the disk cools to a temperature well
below the range of sensitivity of the PCA.  The power-law spectral
index is uniformly harder ($1.86~\leq~\alpha_{pl,LS}~\leq~2.08$) in the
LS than in any observation in the VHS or HS.  The normalization of the
power-law component decreases throughout the LS, as does the overall
flux.  The power-law index measured with Pexriv hardens steadily,
slightly increasing in the last two observations (see Table 2, Figure
4b).

The multicolor blackbody disk model yields an inner accretion disk
radius, with component normalization, source distance, and disk
inclination as parameters.  Fits with this model indicate that the
inner disk radius is relatively constant at $\sim$10~km in the VHS,
and $\sim$19~km for every observation made in the HS (see Table~4).

\subsection{Line Components}
\label{res:line}
Figures 3a and 3b show data/model ratios for the VHS and HS, and LS,
respectively.  The ratios are off-set from each other, but on the same
relative scale.  The models used here are the multicolor blackbody
disk plus Pexriv for the VHS and HS, and Pexriv only for the LS.
These figures serve to illustrate how the equivalent width, width, and
centroid energy of the line features detailed in Table 3 vary across
the outburst.

\placefigure{3}

The bottom data/model ratio plot in Figure 3b corresponds to
observation 23, just before the source slipped beneath RXTE detection
limits.  The line feature is very strong, broadened, and
single-peaked.  Tracing backwards to the beginning of the LS (top of
Figure 3b), it is readily apparent that the line centroid energy is
approximately constant, and the equivalent width is diminished in
earlier observations.  The width of the line feature is measured to be
$\sim$0.30 keV throughout, indicating that the line is resolved
by the PCA in the LS (see Table 3).

The resolved line in the LS allows for tightly-constrained
measurements via fits with the Diskline model.  Centroid energy values
fall within $\pm$0.1 keV of 6.7 keV, and the line equivalent width
increases steadily from $\sim200\pm20$ eV to $\sim916\pm33$ eV.  These
measurements are in good agreement with the values obtained by fitting
the line with a Gaussian profile.  Most notably, the Diskline model
measures the inner edge of the accretion disk, which remains
relatively close to the innermost radius in the LS.  Our fits indicate
the inner edge of the disk is moving closer to the BH in observations
16, 17, and 18, from $63^{+933}_{-31}~R_{\rm g}$ to
$\sim6^{+15}_{0}~R_{\rm g}$.  The inner edge then recedes back to
$102^{+140}_{63}~R_{\rm g}$ in observation 19.  Following this sharp
recession, we measure the inner disk radius to move steadily inward in
subsequent observations, to $\sim35^{+34}_{-15}~R_{\rm g}$ in observation
23.

Fits to the LS are improved by adding a higher-energy Gaussian line
component to the spectral model at $\sim$8.3 keV.  This unresolved
line feature is suggested at $>$68\% confidence throughout the LS, and
is consistent with fluorescent Ni~K$_{\alpha}$ emission from a
highly-ionized species.  The equivalent width of this line varies
between 10--260 eV from observation 16 to observation 23.  The
presence of this line feature is visible in Figure 2b once the more
intense line at $\sim$6.7 keV has been fit, but is not visible in the
plots in Figure 3b as it is far less intense than the lower energy
line.

Following the data/model ratio from the beginning of the LS (top of
Figure 3b) to the end of the HS (bottom of Figure 3a), the $\sim$6.7
keV line feature is still very prominent, and can in fact be traced
all the way back to the VHS (top of Figure 3a).  Although the line
feature is much weaker in the VHS than in the HS or LS, the constancy
of the centroid energy and smooth evolution of the equivalent width
clearly tie the line emission feature near $\sim$6.7 keV in the HS and
VHS to that in the LS.

A lower-energy line feature is present in the HS that is not seen in
the LS.  We distinguish these HS and VHS features as the ``red'' and
``blue'' lines.  Fitting a model including two Gaussian lines to the
HS, we find that the blue line feature is significant at greater than
90\% confidence in all but one observation in the HS (obs.\ 10).  The
centroid energy of the blue line slowly increases from 6.4 to 6.7 keV
from the beginning to the end of the HS.  The equivalent width also
increases steadily, from 39 to 347 eV (see Table 2, Fig. 4a).  The red
line is not significant at 90\% confidence in obs.\ 10, but is measured
in the three observations at the beginning of the HS.

Again tracing backwards in time (bottom to top) in Figure 3a, the
rising equivalent width of the red component and the smeared profiles
of both red and blue line features is clearly seen.  In the VHS the
line emission feature may be characterized as smeared and possibly
double-peaked.  Fits to the VHS spectra with a model including two
Gaussian line emission features find the red line is significant at
$>$99.9\% confidence for every observation.  The blue line is
significant at $>$99.9\% in all but obs.\ 6.  Generally, the equivalent
width and flux of the red line are greater than that of the blue line
in the VHS.

The red line centroid energy decreases slightly through the VHS, but
is consistent with 4.6~keV (see Fig. 4a).  The centroid energy of the
blue line varies in the VHS, steadily increasing from 5.8 to 6.6 keV
(see Table 3).  The red line width is inconsistent with zero in the
VHS except in observations 3 and 6; the blue line except in
observation 7.  The red line is likely at the limit of the PCA
resolution in the VHS; it is unresolved in the HS.  The blue line is
especially broad (0.8~keV) at the beginning of the VHS, becoming
narrower throughout (to 0.4~keV in obs.\ 8).

Fits with the Laor relativistically-skewed line emission model show
that the addition of a line feature emitted from a region between
$1.2~R_{\rm g}$ and $30.0~R_{\rm g}$, with centroid energy of 5.7~keV, is
significant at 90\% confidence in observations 3, 4, and 8.  Later
observations cannot be fit as well with the Laor model.  Indeed, we do
not expect that the resolution of the PCA is sufficient to constrain
Laor model parameters; fits with this model serve primarily as a
consistency check.

The continuity of the changes in line parameters and profiles as we
search backwards from the LS to the VHS indicates that the line-like
excesses seen in Figures 3a and 3b are very likely manifestations of
the same line emission feature.

Tracing the line feature(s) backwards in time from the LS, the
equivalent width is observed to steadily decrease.  It is important to
note that the flux in the red line feature in the VHS is greater than
the flux in the single-peaked/blue line feature in the LS (see Table
3).  Considering the states separately, the flux of the lines is
highest at the beginning of each spectral state, and gradually
decreases throughout (see Table 3, Fig. 4).  More generally, the lines
have the highest flux in the VHS at the beginning of the outburst.
The line profile is possibly double-peaked in the VHS.  As the
outburst moves into the HS, the line profile becomes single-peaked and
increases in equivalent width, though the line flux decreases
steadily.  Finally, as the outburst moves into the LS, the line
equivalent width becomes very large (the soft blackbody disk continuum
component disappears) and the line flux continues to decrease
steadily.

Revnivtsev, Trudolyubov, \& Borozdin (1999) only find evidence for a
line component in the LS spectrum of \xtej, and suggest that the line
is from diffuse galactic emission as the line flux is stable relative
to the decline in the power-law flux.  We disagree with this
interpretation.  Our finding that the line flux is highest at the
beginning of each spectral state and decays until the end of the state
demands that the spectral lines are produced by \xtej.

\subsection{Reflection}
The power-law continuum component Pexriv models the reflection of hot
X-rays impinging on an ionized, optically-thick gas, often assumed to
be an accretion disk.  In the VHS, the reflection fraction $f$ is
small but non-zero at $>$90\% confidence (except obs.\ 6; see Table 2,
Fig. 4a).  Measurements indicate that the disk is very ionized at the
beginning of the VHS ($\xi~\sim~7000$ in obs.\ 3).  The ionization is
constrained only in observations 3, 4, and 5.  Thereafter in the VHS,
and throughout the LS, the ionization is not tightly constrained but a
highly-ionized disk is preferred at 90\% confidence.  We therefore
fixed the ionization ($\xi~=~2000$) in fitting observations 6--15.

\placefigure{4}

Across the VHS--HS transition, $f$ varies discontinuously from
$\sim0.04$ in observation 8 to $\sim0.01$ in observation 9.
Throughout the HS, $f$ increases to a value of $\sim0.11$ in
obs.\ 15.  Perhaps due to the dominance of the soft component in the
HS, $f$ is consistent with no reflection throughout.  

The reflection fraction $f$ again varies sharply across the HS--LS
transition, from $\sim0.11$ in observation 15 to $0.45\pm0.02$ in
observation 17.  The value of $f$ falls sharply in subsequent
observations, to $\sim0.04$ in observation 21, and becomes consistent
with zero in obs.\ 23 (see Table 2, Fig. 4b).  Measurements of $\xi$ in
the LS are consistent with a neutral disk, marking a sharp contrast to
the highly-ionized disk required in the VHS and HS.

\section{Discussion}

\subsection{The Fe~K$_{\alpha}$ Fluorescent Line}
Based on the broadened, redshifted line emission profile observed by
Iwasawa \etal\ in MCG$-$6-30-15 (1997), and on the profiles and
equivalent widths calculated by many researchers (e.g. Magdziarz \&
Zdziarski 1995, Fabian \etal\ 1989, George \& Fabian 1991, Dabrowski
\etal\ 1997, Martocchia \etal\ 2000), we interpret the red and blue
line features in the VHS of \xtej\ as wings of a Doppler- and
gravitationally-shifted Fe~K$_{\alpha}$ line emission profile.  The
line profiles (Figs.~2a and 3a; Table~3) are consistent with those
calculated by Martocchia \etal\ (2000) for emission from the innermost
region of an accretion disk orbiting a BH of high angular momentum,
viewed at a modest inclination angle, with a source of hot X-rays
located $\leq~10~R_{\rm g}$ above the BH.  The equivalent widths measured
in the VHS are broadly consistent with those predicted by the model
for a disk with outer emission radius near $100~R_{\rm g}$, and viewed at
an angle slightly below the 45 degree inclination we assumed.

Our fits to the VHS using the Laor model indicate that in this
spectral state the relativistically-skewed line is consistent with
emission from an inner radius extending down to the marginally stable
orbit for a maximally rotating Kerr black hole, $1.2~R_{\rm g}$ (see
Table~4), in obs.\ 3 and obs.\ 4.  Measurements of the inner disk
radius via the multicolor blackbody disk model in the VHS confirm that
the disk may extend to the innermost stable orbit in a Kerr geometry
(see Table~4).

Throughout the HS, the red wing of the profile becomes less prominent,
and the blue wing strengthens.  In particular, the centroid energy of
the blue line shifts from 5.8~keV (beginning of the VHS) to 6.7~keV
(end of the HS).  Throughout the LS, the line profile is
single-peaked, and measured at 6.70$\pm$0.10~keV.  This value is
consistent with highly ionized species of iron.  We suggest that this
line is likely a complex of lithium-like, helium-like, and hydrogenic
iron species (Fe~XXIV-XXVI).  The shift of line profile to higher
centroid energies may be produced in two ways, either by an evolving
ratio of iron ionizations contributing to the line profile, or by the
disk mildly recessing from the BH.  Given the inner disk radii we
measure (see Table~2) across the outburst and the high ionization
preferred in the VHS and HS, it is likely that the latter process
dominates, but both may contribute.

Although unlikely given the relatively continuous evolution of the
line profile (in contrast to the discontinuous nature of the jet,
which is only reported in the VHS), it is nevertheless possible that
the lines originate within the jet itself.  It is also possible that
the lines are produced within a hot coronal region, and the measured
reflection is from a cold geometry apart from the disk.

\subsection{Disk Reflection and Ionization}

The reflection fractions we measure in the VHS and HS are smaller than
the same parameter measured in observations of other BHXBs and AGN.
Done, Madejski, \& Zycki (2000) measure $f$ between $0.55-1.25$ in
Seyfert 1 galaxy IC4329a.  Zycki, Done, \& Smith (1998) measure
$f~\sim~0.30$ in the VHS, and $f~\sim~0.64$ in the HS of the 1991
outburst of GS~1124$-$68.  Zycki \etal\ (1999) measure
$f~\sim~0.4-0.9$ during the decline of GS~2023$+$338 (this source was
dominated by the hard power-law component through its decline, unlike
\xtej\ and GS~1124$-$68).  The reflection fraction measured in the
early LS observations of \xtej\ is comparable to that measured in the
LS of Cygnus X-1 by Gierlinski \etal\ (1999)
$f_{cyg,LS}~\sim~0.5-0.7$.

The low reflection fractions we measure are likely due to the high
disk ionizations strongly preferred in the VHS and HS.  The
contrasting high reflection and negligible ionization at the beginning
of the LS is not surprising; a similar phenomenology is noted by
Zycki, Done, \& Smith (1998) for the HS-LS transition in GS~1124$-$68.
In a geometry where the disk is illuminated by a central source of hot
X-rays, the disk should be most ionized when the hard X-ray flux is
highest and when the disk is closest to the central source.  The hard
X-ray fluxes and inner disk radii measured in \xtej\ support this
picture.  Moreover, the lines that we measure in the VHS and HS,
though often significant at 99.9\% confidence, are still weaker than
those seen in AGN spectra.  Therefore, the strength of the two
components predicted for reflection spectra -- fluorescent line
production and disk reflection, are at least qualitatively consistent.

Ross, Fabian, and Young (1999) note that even the small departures
from uniformity that may occur in a highly ionized disk can account
for the small reflection fraction often observed is such scenarios,
without invoking reflection from the outer disk.  This research, and
that of Nayakshin, Kazanas, \& Kallman (1999) finds that the hottest
outer layers of the accretion disk are important in accurately
determining disk reflection fractions.  These studies find that Pexriv
may be too basic in its treatment of Compton processes in these hot
outer layers.  At present, Pexriv is the best code publicly available.
We look forward to the release of more sophisticated codes and to high
resolution data from Chandra and XMM-Newton to constrain new models.

We also note that we are only able to obtain measurements for the
reflection fraction inconsistent with zero when the hard flux
dominates the 3.5-20.0 keV bandpass we fit.  This condition is met
throughout the VHS, in obs.\ 15 in the HS, and throughout the LS.
With the exception of obs.\ 15, we are not able to place tight
constraints on reflection in the HS.  Many models for disk reflection
are based upon AGN spectra (the codes we mention above are notable
exceptions), which generally do not contain a soft component in the
X-ray band.  The extent to which the physical process of soft X-ray
emission from the disk is responsible for confusing disk reflection
measurements, relative to merely poor mathematical modeling, is
unknown.

\subsection{Implications for Geometry and Accretion Flow}

DiMatteo \& Psaltis (1999) establish a possible connection between
QPO frequency and inner disk radius in BHXBs based on similar work on
neutron star (NS) systems:
\begin{equation} 
  \left(\frac{R_{\rm in}}{R_{\rm g}}\right) \leq 27~\nu_{1}^{-0.35}
  \left({\frac{M}{2\Msun}}\right)^{-2/3},
\end{equation}
where $\nu_{1}$ is the QPO frequency, $R_{\rm in}$ is the inner disk
radius, $R_{\rm g}$ the gravitational radius, $M$ the mass of the black
hole, and $\Msun$ a solar mass.  For a measured QPO frequency, and a
measured or assumed black hole mass, the inner radius can be
determined via this equation.  Inserting the 32 Hz QPO (Fox \& Lewin,
1998) measured in the VHS of \xtej, and assuming a black hole mass of
10\Msun, this equation predicts an upper limit for the accretion disk
inner radius of $\sim2.75~R_{\rm g}$.  This value supports the
measurements we obtain via the Laor line model, and our comparison to
the models of Martocchia \etal\ (2000).

The multicolor blackbody disk continuum model also provides a measure
of the inner disk radius.  A central black hole with a mass of
10$\Msun$ has a gravitational radius of $\sim$15~km; a central black
hole with mass 5$\Msun$ a gravitational radius of $\sim$7.5~km.  Fits
using the multicolor blackbody disk model measure the inner disk
radius in the VHS to be $\sim$10~km.  Fits with the Laor disk line
emission model in the VHS are consistent with an inner disk radius
extending down to 1.2$R_{\rm g}$; this corresponds to $\sim$18~km for
$M_{BH}=10\Msun$, and $\sim$9~km for $M_{BH}=5\Msun$.  Finally, the
connection noted by DiMatteo \& Psaltis (1999) suggests an inner disk
radius of $\sim2.75R_{\rm g}$, or 41.3~km and 20.6~km for 10$\Msun$
and 5$\Msun$ black holes, respectively.  Although these methods do not
yield exactly the same inner disk radii, they all indicate that the
inner edge of the accretion disk extends below the marginally stable
orbit for a non-rotating Schwarzschild black hole ($R_{\rm ms,Sch} =
6R_{\rm g} = 90$~km; for $M_{BH} = 10\Msun$, $R_{\rm ms,Sch} = 6R_{\rm
g} = 45$~km, for $M_{BH} = 5\Msun$; see Table~4 for $R_{\rm in}$
results).

Narayan and Yi (1994) describe a solution for BH accretion geometry
wherein a hot, optically thin, inefficiently-cooled plasma advects
energy directly onto the BH.  Originally developed to describe the
presence of hard emission within the spectra of AGN, these
Advection-Dominated Accretion Flow (ADAF) models successfully describe
a number of AGN (see Narayan 1997 for a review).  A natural
application of this model is soft X-ray transients in quiescence,
where $\dot{m}$ is sufficiently small.  BHXBs V404 Cygnus and A0620-00
are well fit by ADAF models in quiescence (Narayan, McClintock, \& Yi 1996;
Narayan, Barret, \& McClintock 1997a).

Esin, McClintock, and Narayan (1997) describe an ADAF/thin-disk
composite model tuned to explain the full spectral state evolution of
BHXBs over a range of luminosities and $\dot{m}$.  The parameter which
drives the spectral transitions is the location of the ADAF/disk
transition radius, $R_{tr}$.  When $R_{tr}$ is close to the BH, the
emission is dominated by the disk.  As the outburst evolves, cooling
becomes less efficient, $\dot{m}$ falls, the inner region becomes an
ADAF, and $R_{tr}$ increases.  Esin \etal\ predict that in the VHS and
HS, the disk may extend to the innermost stable orbit, but in the LS
and QS $R_{tr}$ is likely to be near $\sim~10^{4}~R_{\rm g}$.  This model
is applied to the 1991 outburst of GS~1124$-$68, and describes all but
the VHS.  Subsequent application of the model to Cygnus X-1 in its LS
indicates $R_{tr}\sim100~R_{\rm g}$ (Esin \etal\ 1998).

Zycki, Done, \& Smith (1998) constrain the hot advection zone to lie
within $\sim20-100~R_{\rm g}$ in the LS of GS~1124$-$68, based on smearing
of the Compton upscattering excess observed above $\sim$10 keV.  The
inner radii we measure in the LS of \xtej\ via the multicolor
blackbody disk model and the Diskline relativistic line model also
constrain the inner disk edge to lie in this range.  We therefore
conclude that a large ($10^{4}~R_{\rm g}$) inner advection region of the
kind predicted by Esin, McClintock, and Narayan is not consistent with
the LS of \xtej\, but must be within 100~$R_{\rm g}$ (similar to Cygnus
X-1) if such a geometry exists.

We make note of work by Merloni, Fabian, \& Ross (2000), which finds
that the multicolor blackbody disk model systematically underestimates
the real inner disk radius, and can suggest inner disk motion at high
accretion rates when in fact the disk is stable at the marginally
stable orbit.  In the VHS, we find that the multicolor blackbody disk
radius is stable and corresponds to the marginally stable orbit for a
Kerr black hole (see Table~4).  The radii we measure via this model in
the VHS are supported by radii derived via DiMatteo \& Psaltis (1999),
but may be an important caveat for HS measurements.

\subsection{Constraints on the BH Mass}

Combining the relation discussed by DiMatteo \& Psaltis with the
multicolor blackbody disk normalization in the VHS and the radio-band
distance estimate, we obtain the following constraint for the mass of
the central object in \xtej:
\begin{equation} 
  M \geq 7.9\Msun
  \left(\frac{D}{8~\mbox{kpc}}\right)^3\left(\frac{\cos \theta}{\cos 45^{\circ}}\right)^{-3/2}\left(\frac{N}{100}\right)^{3/2}\left(\frac{\nu}{32~\mbox{Hz}}\right)^{1.05},
\end{equation}
where $M$ is the mass of the compact object, $\Msun$ the solar mass,
$D$ the distance in kiloparsecs, $N$ is the normalization of the
Diskbb component (see Table~2), $\theta$ the disk inclination angle,
and $\nu$ the QPO frequency.  Although this calculation is subject to
numerous uncertainties, including the uncertain color correction
(Ebisawa \etal\ 1994), taken at face value it implies a black hole
primary ($M > 4.5\Msun$) in \xtej\ for a large parameter space
including all likely values of $D$ ($>$8~kpc from radio measurements)
and $\cos\theta$ ($<$1).

\subsection{Comparison to Recent Evidence of Lines in BHXBs}
Cui, Chen, \& Zhang (1999) find two co-moving line features placed
quasi-symmetrically about 6.5 keV in the 1996 outburst of 4U~1630-47.
It is suggested that each may be due to Doppler-shifted
Fe~K$_{\alpha}$ emission (not principally gravitational redshifting,
in the case of the reddened line).  

Balucinska-Church \& Church (2000) report on four spectra from
GRO~J1655$-$40, and find evidence for gravitationally redshifted and
Doppler-broadened emission.  Our finding that the red emission wing is
more prominent than the blue emission wing early in the VHS of \xtej\
is qualitatively similar to these results.  Additionally, fits with
Pexriv to these observations of GRO~J1655$-$40 find limits for disk
reflection ($f~<$~1\%) and ionization ($\xi~<~10^{4}$).  Like the HS of
\xtej\, these observations occur when the flux is dominated by the
soft disk component, likely complicating efforts to measure any
reflection.

Unfortunately, 4U~1630$-$47 and GRO~J1655$-$40 are both known to be
``dipping'' sources (Kuulkers \etal\ 1988) -- to experience periods of
markedly different absorption during outburst -- and this complicates
efforts to trace spectral lines across a full outburst and thereby
constrain accretion flow geometries.  

\section{Conclusions}
We have analyzed the 22 publicly-available observations between
maximum X-ray luminosity and quiescence from the outburst of galactic
microquasar BHXB \xtej.  The spectra can be categorized into Very
High, High, and Low States based on the relative flux contributions
from hard and soft continuum components.  We trace emission lines
through these spectral states, and find evidence for an
Fe~K$_{\alpha}$ line that is gravitationally redshifted and Doppler
shifted in the VHS, with a profile that is consistent with emission
from the innermost stable orbit for a maximally rotating Kerr black
hole.  The line is less redshifted but increasingly prominent in the
HS.  Finally, in the LS, the line is centered at 6.7$\pm$0.10~keV, and
we measure a profile that is consistent with emission from a disk with
inner radius between $R\sim20$ and $R\sim100~R_{\rm g}$.

We fit each observation with a power-law continuum model (Pexriv) that
includes Compton reflection of hot incident X-rays from an ionized
disk component, manifested spectrally as a broad excess above $\sim$10
keV.  This work represents the first trace of the disk reflection
fraction across the full outburst of a BHXB.  The fractions we measure
in the VHS and HS are much smaller than those often measured in AGN,
but well-constrained.  We note that the Pexriv model may be too basic
for accurate reflection measurements, and urge the public release of
new codes.

The Fe~K$_{\alpha}$ line serves as a direct diagnostic of the
accretion geometry.  We consider our findings in the context of the
ADAF model developed by Esin, McClintock, \& Narayan (1997) to
describe the spectral states of BHXBs.  As the disk in \xtej\ is
measured to be a few tens of $R_{\rm g}$ from the BH in the LS, rather
than $10^{3}$-$10^{4}~R_{\rm g}$, we find that the predictions of this
ADAF model are inconsistent with the accretion geometry of \xtej.

We look forward to data from Chandra and XMM to obtain
higher-resolution spectra of a BHXB in outburst.

\section{Acknowledgments}
Special thanks is due Taro Kotani for his communication of ASCA
$N_{\rm H}$ measurements in advance of publication.

We thank Michael Muno, Dimitrios Psaltis, Rob Fender, Keith Jahoda,
Andrzej Zdziarski, and Wei Cui for numerous insights .  We are
especially indebted to Ron Remillard for his many helpful discussions.

We also wish to acknowledge the referee, who provided many helpful
comments.

WHGL gratefully acknowledges support from NASA.  TDM and RW were
supported by NASA through Chandra Fellowship grant numbers PF9-10005
and PF-9-10010, respectively, which are operated by the Smithsonian
Astrophysical Observatory for NASA under contract NAS8-39073.  This
research has made use of data and resources obtained through the
HEASARC Online Service, provided by the NASA-GSFC.

\clearpage

\begin{table}[t]
\caption{RXTE PCA Observation Log}
  \scriptsize
  \begin{center}
  \begin{tabular}{lrllllll}
No. &  Obs.ID & Date, UT & Start Time & PCA Exp. (s)\\ 
\tableline
\tableline
1 & 30188-05-01-00 & 04/06/98 & 20:05:04 & 1760 \\ 
2 & 30188-05-02-00 & 05/06/98 & 03:03:44 & 768 \\ 
3 & 30171-02-01-00 & 06/06/98 & 09:41:20 & 2655 \\
4 & 30185-01-01-00 & 07/06/98 & 07:56:32 & 2944 \\
5 &      ...-02-00 & 08/06/98 & 06:23:28 & 3027 \\
6 &      ...-03-00 & 09/06/98 & 12:48:00 & 3729 \\
7 &      ...-04-00 & 10/06/98 & 03:38:24 & 7721 \\
8 &      ...-05-00 & 11/06/98 & 12:52:16 & 3439 \\
9 &      ...-06-00 & 13/06/98 & 12:51:28 & 3114 \\
10 &     ...-07-00 & 15/06/98 & 04:53:36 & 1795 \\
11 &     ...-08-00 & 18/06/98 & 20:55:12 & 2327 \\
12 &     ...-09-00 & 22/06/98 & 22:30:08 & 3210 \\
13 &     ...-10-00 & 27/06/98 & 11:39:28 & 1647 \\
13.1 &   ...-10-01 & 02/07/98 & 18:18:24 & 1280 \\
14 &     ...-11-00 & 08/07/98 & 16:21:52 & 1295 \\
15 &     ...-12-00 & 13/07/98 & 06:44:00 & 2056 \\
16 &     ...-13-00 & 18/07/98 & 04:00:32 & 10585 \\
17 &     ...-14-00 & 30/07/98 & 09:44:48 & 6841 \\
18 &     ...-15-00 & 05/08/98 & 18:25:20 & 4333 \\
19 &     ...-16-00 & 13/08/98 & 10:17:36 & 1565 \\
20 &     ...-17-00 & 20/08/98 & 16:41:20 & 1704 \\
21 &     ...-18-00 & 25/08/98 & 03:32:00 & 1785 \\
22 &     ...-19-00 & 14/09/98 & 08:17:36 & 886 \\
23 &     ...-20-00 & 26/09/98 & 03:29:04 & 10287 \\
\tableline
\tableline
\end{tabular} ~\vspace*{\baselineskip}~\\ \end{center}
\tablecomments{Observations 1 and 2 are during
  the rise of the outburst.  The remaining observations were made
  during the declining portion of the outburst, and are public data.}
  \vspace{-1.0\baselineskip}
\end{table}

\clearpage

\begin{figure}
\figurenum{1}
\label{fig:Crab with pcarsp response file}
\centerline{~\psfig{file=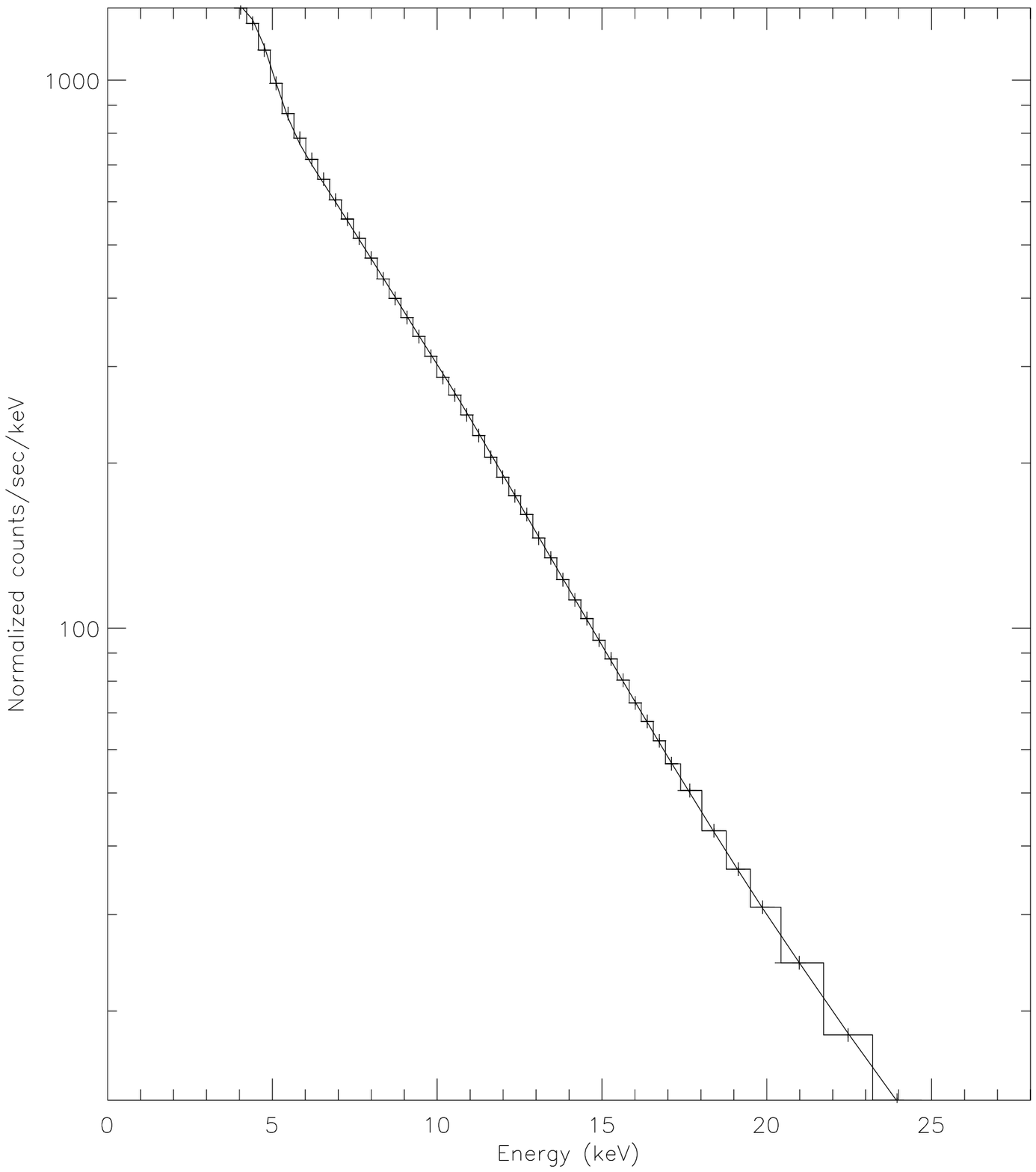,width=6.0in,height=3.5in}~}
\centerline{~\psfig{file=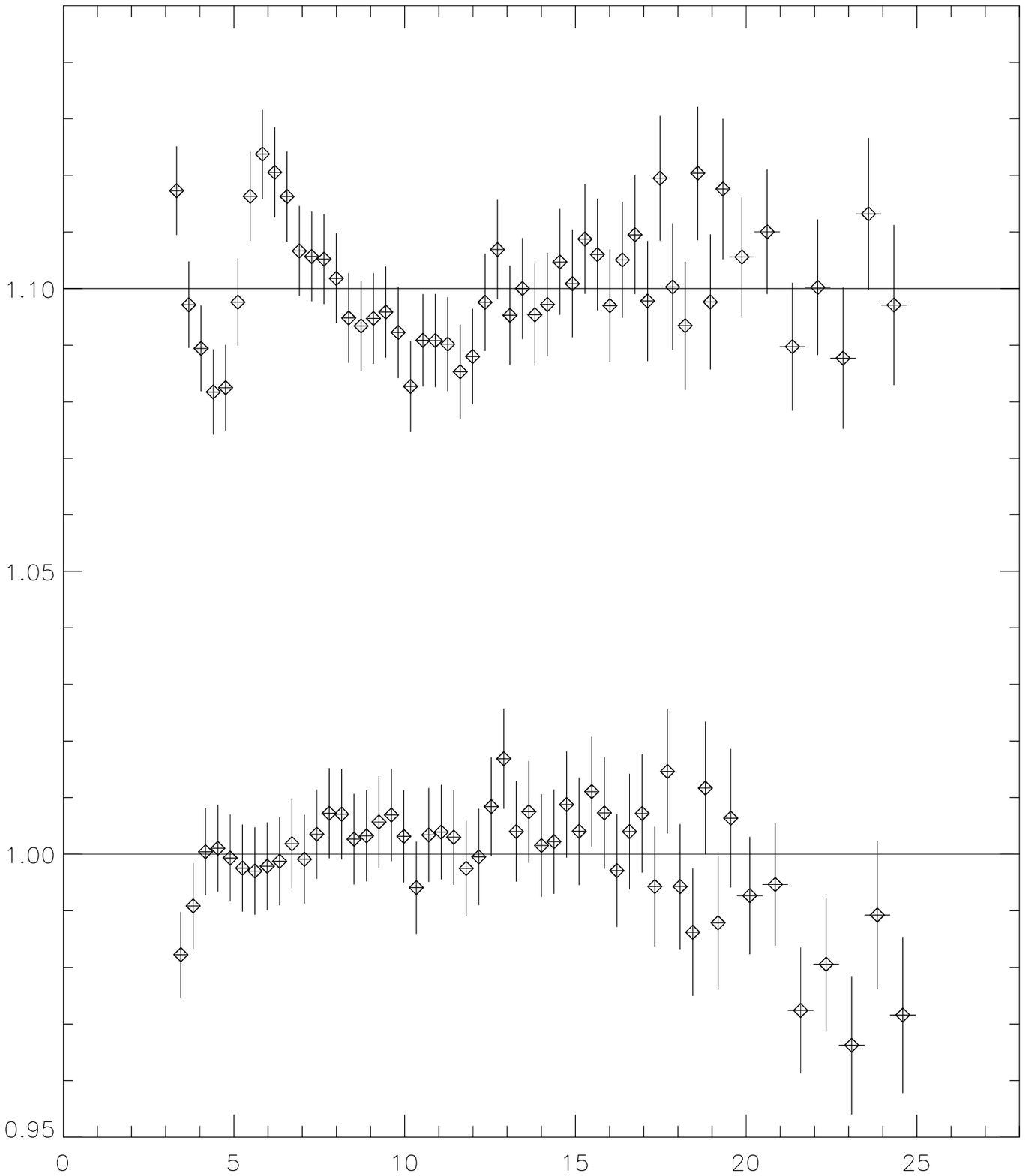,width=6.0in,height=3.5in}~}
\caption{Above: May 1998 Crab Nebula spectrum, fit with power-law and
low-energy absorption model.  Below: May 1998 Crab data/model ratios,
pcarsp (top) and static (bottom) rsp files.}
\end{figure}

\clearpage

\begin{table}[t]
\caption{Spectral Component Parameters of \xtej}
  \tiny
  \begin{center}
  \begin{tabular}{lllllllllllll}
Obs. &  $T_{DBB}$ & Norm. & $R_{DBB}$ & Flux$_{DBB}$ & $\alpha_{pl}$ & $f$ & F-stat. & $\xi$ & Norm. &  Flux$_{pl}$ & Flux$_{tot}$ & $\chi^{2}/dof$\\ 
~ & (keV) & ~ & (km) & $10^{-9}~cgs$ & ~ & $10^{-2}$ & ~ & $10^{3}$ & ~ & $10^{-9}~cgs$  & $10^{-9}~cgs$ & \\
\tableline
\tableline
\multicolumn{13}{c}{---  Very High State  ---}\\
\tableline
3 & $1.57(1)$ & $90(3)$ & 9.0(1) & $3.6(1)$ & 2.728(1) & $2.1(6)$ & 0.079  &$7^{+17}_{-6}$ & $15.59(5)$ & $7.82(2)$ & $11.4(1)$ & 1.02\\
4 & $1.53(1)$ & $104(4)$ & 9.7(1) & $3.7(1)$ & 2.736(1) & $2.2(7)$ & 0.036  &   $3^{+8}_{-2}$ & $15.48(5)$ & $7.63(2)$ & $11.3(1)$ & 0.64 \\
5 & $1.29(2)$ & $108(9)$ & 9.9(3) & $1.5(1)$ & 2.631(1) & $2.2(6)$ & 0.001 &$3^{+4}_{-2}$ & $13.80(3)$ & $8.43(2)$ & $9.9(1)$ & 0.49\\
6 & $1.19(2)$ & $125(13)$ & 10.6(5) & $1.1(1)$ & 2.515(1) & $0.5(4)$ & 0.403  &$2^{\dag}$ & $10.57(3)$ & $8.15(2)$ & $9.3(1)$ & 0.45\\
7 & $1.22(2)$ & $116(11)$ & 10.2(4) & $1.1(1)$ & 2.571(1) & $6.6(6)$ & $<$0.001   &$2^{\dag}$ & $11.07(2)$ & $7.68(2)$ & $8.8(1)$ &0.34\\
8 & $1.26(2)$ & $88(6)$ & 8.9(2) & $1.0(1)$ & 2.466(1) & $3.6(9)$ & $<$0.001 &$2^{\dag}$ & $5.43(2)$ & $4.64(2)$ & $5.6(1)$ &0.51\\
\tableline
\multicolumn{13}{c}{---  High State  ---}\\
\tableline
9 & $1.29(1)$ & $388(6)$ & 18.7(2)& $5.3(1)$ & 2.73(2) & $2^{+5}_{-2}$ & 1.0 &$2^{\dag}$ & $3.62(2)$ & $1.81(1)$ & $7.1(1)$ &0.42\\
10 & $1.31(1)$ & $370(6)$ & 18.3(2) & $5.5(1)$ &  2.94(3) &$2^{+12}_{-2}$ & 1.0    & $2^{\dag}$ & $3.5(2)$ & $1.15(8)$ & $6.7(1)$ &0.69 \\
11 & $1.26(1)$ & $403(6)$ & 19.1(1) & $5.9(1)$ & 2.90(6) &$2^{+4}_{-2}$ & 1.0 &$2^{\dag}$ & $2.1(3)$ & $0.72(8)$ & $6.6(1)$ &0.55\\
12 & $1.20(1)$ & $436(8)$ & 19.9(1) & $3.8(1)$ &  2.69(7) &$1^{+21}_{-1}$ & 1.0 & $2^{\dag}$ & $1.0(1)$ & $0.53(1)$ & $4.3(1)$&0.70 \\
13 & $1.15(1)$ & $440(8)$ & 20.0(1) & $3.3(1)$ &  2.58(1) &$3^{+4}_{-3}$ & 1.0 &$2^{\dag}$ & $0.87(8)$ & $0.57(7)$ & $3.9(1)$ &0.69\\
13.1 & $1.03(1)$ & $450(10)$ & 20.2(3) &$1.6(1)$ &  2.47(5) & $7^{+22}_{-7}$ & 1.0 &$2^{\dag}$ & $0.89(2)$ & $0.68(1)$ & $2.3(1)$ &0.45\\
14 & $0.98(1)$ & $440(10)$ & 20.0(3) & $1.1(1)$ &  2.41(7) & $3^{+26}_{-3}$ & 1.0 & $2^{\dag}$ & $0.63(7)$ & $0.6(1)$ & $1.7(1)$ & 1.13\\
15 & $0.91(1)$ & $390(10)$ & 18.8(3) & $0.56(5)$ &  2.44(5) & $11.1(1)$ & $<$ 0.001 & $2^{\dag}$& $0.97(1)$ & $0.87(2)$ & $1.4(1)$ &0.94\\
\tableline
\multicolumn{13}{c}{---  Low State  ---}\\
\tableline
16 & $0.42(3)$ & $7800^{7500}_{3700}$ & $84(36)$ & $0.02(1)$ & 1.86(2)& $8^{+4}_{-2}$ & $<$0.001 & $0$ & 0.32(1) & $1(1)$ & $1.1(1)$ & 0.75\\
17 & $0.34(9)$ & $12900^{164100}_{5200}$ & $100^{300}_{20}$ & $0.03(1)$ & 1.98(1) & 45(2) & $<$0.001 & $0$ & 0.24(1) & $0.64(1)$ & $0.67(1)$ &0.62\\
18 & $0.34(8)$ & $14400^{173000}_{7200}$ & $100^{300}_{20}$ & $0.02(1)$ & 1.95(2) & $21(2)$ & $<$0.001 & $0$ & 0.20(1) & $0.54(2)$ & $0.57(1)$ & 1.02 \\
19 & -- & -- & -- & -- & 2.02(1) & $28(3)$ & $<$0.001 & $0$ & 0.21(1) & $0.49(2)$ & $0.51(1)$ &0.85\\
20 & -- & -- & -- & -- & 1.90(1) & $4(2)$ & 0.254 & $0.15^{+0.85}_{-0.15}$ & 0.11(1) & $0.32(1)$ & $0.34(1)$&0.78\\
21 & -- & -- & -- & -- & 1.91(1) & $4(3)$ & 1.0 & $0^{+0.6}$ & 0.10(1) & $0.28(1)$ & $0.31(1)$ &0.72\\
22 & -- & -- & -- & -- & 2.17(1) & $11(3)$ & $<$0.001 & $0$ & 0.17(1) & $0.28(1)$ & $0.31(1)$ &0.77\\
23 & -- & -- & -- & -- & 2.08(1) & $1.0(1)$ & 1.0 &$0$ & 0.10(1) & $0.21(1)$ & $0.23(1)$ &0.40\\
\tableline
\tableline
\end{tabular} ~\vspace*{\baselineskip}~\\ \end{center} 
\tablecomments{Basic model parameters and normalizations quoted with
  90\% confidence limits.  Model components include a multicolor
  blackbody, a Gaussian, and Pexriv.  All fluxes are measured in the
  3.5--20.0 keV band, as are all model parameters and normalizations.
  Where errors are not quoted, the error does not affect the value of
  the last significant digit.  Fits to the VHS and HS have 30 $dof$
  when red and blue lines are fit, and 36 $dof$ when lines are not
  fit.  Fits to the LS have 33 $dof$ when the blue line is fit, 36
  $dof$ when they are not.  Quoted $\chi^{2}$ values are for models
  including lines.  All observations fit with 0.75\% systematic errors
  added, and (fixed) $N_{H} = 7.5\times10^{22}~cm^{-2}$.  $R_{DBB}$ is
  the disk-blackbody-derived inner radius, calculated for a disk
  inclination of 45 degrees (assumed), and a distance of 8~kpc.  The
  Pexriv power-law cut-off energy is fixed at 200~keV, and the disk
  inclination at 45 degrees.  $f$ is the disk reflection fraction,
  $0.0 < f < 1.0$ corresponds to the range $0 - 2\pi$, and F-stat is
  the standard F statistic, measuring the significance of $f$.\\
  $^{\dag}$ denotes an observation where the ionization is fixed at
  2000.}  \vspace{-1.0\baselineskip}
\end{table}

\clearpage

\begin{table}[t]
\caption{Fe~K$_{\alpha}$Line Parameters}
  \scriptsize
  \begin{center}
  \begin{tabular}{lllllllll}
\tableline
\tableline
Obs. & Centroid Energy & Width & Eq.Width &
  Flux & $\chi^{2}$/dof & $\chi^{2}$/dof & $F$-stat.\\ 
~ & (keV) & (keV) & (eV) & $10^{-11}~cgs$ & w/o line & w/ line & ~  \\
\tableline
\tableline
\multicolumn{8}{c}{---  Red Wing  ---}\\
\tableline
\tableline
\multicolumn{8}{c}{---  Very High State  ---}\\
\tableline
3 & $4.6(1)$ & $0.2^{+0.1}_{-0.2}$ & $76(15)$ & $17(2)$ & 4.19 & 1.02 & $<$0.001\\
4 & $4.6(1)$ & $0.4(1)$ & $105(15)$ & $22(3)$ & 5.62 & 0.64 & $<$0.001\\
5 & $4.6(3)$ & $0.3(2)$ & $69(14)$ & $12(3)$ & 2.97 & 0.49 & $<$0.001\\
6 & $4.5(2)$ & $0.3^{+0.2}_{-0.3}$ & $51(14)$ & $8(2)$ & 1.87 & 0.45 & $<$0.001\\
7 & $4.5(2)$ & $0.3(3$)& $57(15)$ & $8(2)$ & 1.96 & 0.34 & $<$0.001\\
8 & $4.7(1)$ & $0.5(1)$ & $96(14)$ & $9(1)$ & 4.62 & 0.50 & $<$0.001\\
\tableline
\multicolumn{8}{c}{---  High State  ---}\\
\tableline
9 & $4.5(2)$ & $0.3(3)$ & $39(12)$  & $7(2)$ & 1.35 & 0.42 & $<$0.001\\
10$^{\dag}$ & $4.5(3)$ & $0^{+0.5}$ & $23(9)$  & $4(2)$ & 1.00 & 0.69 & 0.14\\
11 & $4.5(2)$ & $0^{+0.5}$ & $20(9)$  & $3(1)$ & 0.83 & 0.45 & 0.04\\
12 & -- & -- & -- & -- & -- & -- & --\\
13 & -- & -- & -- & -- & -- & -- & --\\
13.1 & -- & -- & -- & -- & -- & -- & --\\
14 & -- & -- & -- & -- & -- & -- & --\\
15 & -- & -- & -- & -- & -- & -- & --\\
\tableline
\tableline
\\
\tableline
\tableline
\multicolumn{8}{c}{---  Blue Wing  ---}\\
\tableline
\tableline
\multicolumn{8}{c}{--- Very High State  ---}\\
\tableline
3 & $5.8(1)$ & $0.8(1)$ & $120(20)$ & $19(4)$ & 5.42 & 1.02 & $<$0.001\\
4 & $6.3(2)$ & $0.4(2)$ & $66(16)$  & $10(2)$ & 2.49 & 0.64 & $<$0.001\\
5 & $6.3(3)$ & $0.5(3)$ & $51(19)$  & $6(2)$ &  1.57 & 0.49 & $<$0.001\\
6 & $6.4(3)$ & $0.5(4)$ & $45(16)$ &  $5(2)$ &  1.25 & 0.45 & 0.002\\
7 & $6.4(3)$ & $0.4(4)$ & $39(18)$ &  $4(1)$ &  1.09 & 0.34 & $<$0.001\\
8 & $6.6(2)$ & $0.4(3)$ & $60(18)$  &  $4(1)$ & 2.00 & 0.50 & $<$0.001\\
\tableline
\multicolumn{8}{c}{--- High State  ---}\\
\tableline
9  & $6.4(3)$ & $0.4^{+0.3}_{-0.4}$ & $39^{+19}_{-12}$  & $4(2)$ & 1.20 & 0.42 & 0.002\\
10$^{\dag}$ & $6.5(3)$ & $0.3^{+0.6}_{-0.3}$ & $30^{+20}_{-12}$  & $3(1)$ & 1.07 & 0.69 & 0.103\\
11 & $6.6(3)$ & $0.1^{+0.7}_{-0.1}$ & $29^{+15}_{-12}$ & $2(1)$ & 1.03 & 0.55 & 0.036\\
12 & $6.8(2)$ & $0.0^{+0.5}$ &  $33(10)$ & $3(1)$ & 1.31 & 0.70 & 0.036\\
13 & $6.7(1)$ & $0.0^{+0.3}$ &  $56^{+14}_{-11}$ & $2(1)$ & 2.19 & 0.69 & $<$0.001\\
13.1 & $6.8(1)$ & $0.2(2)$ &  $88^{+19}_{-16}$ & $2.0(4)$ & 3.13 & 0.45 & $<$0.001\\
14 & $6.74(9)$ & $0.2(2)$ &  $130(19)$ & $2.0(3)$ & 5.96 & 1.13 & $<$0.000\\
15 & $6.69^{+0.09}_{-0.04}$ & $0.2(1)$ & $347(31)$ & $4.8(5)$ & 23.7 & 0.94 & $<$0.001\\
\tableline
\multicolumn{8}{c}{--- Low State  ---}\\
\tableline
16 & $6.68(4)$ & $0.06^{+0.14}_{-0.06}$ & $193(13)$ & $1.8(1)$ & 16.9 & 0.75 & $<$0.001\\
17 & $6.72(3)$ & $0.30(5)$ & $420(19)$ &  $2.3(1)$ & 44.6 & 0.62 & $<$0.001\\
18 & $6.71(3)$ & $0.32(4)$ & $508(22)$ &  $2.4(1)$ & 50.1 & 1.02 & $<$0.001\\
19 & $6.68(3)$ & $0.20(8)$ & $545(25)$ &  $2.3(1)$ & 32.6 & 0.85 & $<$0.001\\
20 & $6.80(3)$ & $0.24(7)$ & $574(30)$ &  $1.6(1)$ & 27.0 & 0.78 & $<$0.001\\
21 & $6.77(3)$ & $0.24(6)$ & $661(36)$ &  $1.7(1)$ & 25.8 & 0.72 & $<$0.001\\
22 & $6.65(4)$ & $0.34^{+16}_{-12}$ & $931(54)$ & $2.5(1)$ & 30.4 & 0.77 & $<$0.001\\
23 & $6.78(3)$ & $0.26(4)$ & $967(40)$ &  $1.6(1)$ & 30.5 & 0.40 & $<$0.001\\
\tableline
\tableline
\end{tabular} ~\vspace*{\baselineskip}~\\ \end{center} 
\tablecomments{Line parameters, normalizations, and fluxes, are all
  quoted with 90\% confidence limits.  Fits to the VHS and HS have 30
  $dof$ when red and blue lines are fit, and 36 $dof$ when lines are
  not fit.  Fits to the LS have 33 $dof$ when the blue line is fit, 36
  $dof$ when they are not. The $F$-statistic is the classic f-test
  statistic.  Observations 12, 13, 13.1, 14, and 15 are not consistent
  with the inclusion of a red wing.\\ $\dag$ denotes observations
  where measurements constrain parameters but do not meet our
  criterion for statistical significance.}
\vspace{-1.0\baselineskip} 
\end{table}

\clearpage

\begin{table}[t]
\caption{Three Methods to $R_{\rm in}$ in the VHS}
  \scriptsize
  \begin{center}
  \begin{tabular}{lllll}
Obs. & $\nu_{QPO}^{(a)}$ & $R_{\rm in}^{(b)}$ & $R_{in}^{(c)}$ &
$R_{\rm in}^{(d)}$\\
~ & (Hz) & ($R_{\rm g}$) & ($R_{\rm g}$) & ($R_{\rm g}$) \\
\tableline
\tableline
3 & 31.6(2) & 4.38(1) & 2.8(1) & 1.2(1)\\
4 & 31.3(2) & 4.39(1) & 2.8(1) & 1.3(1)\\
5 & 23.7(1) & 4.84(1) & 3.0(1) & 1.3(1)\\
6 & 20.2(1) & 5.12(1) & 3.2(1) & 1.4(1)\\
7 & 20.0(1) & 5.14(1) & 3.2(1) & 1.4(1)\\
8 & 22.6(2) & 4.92(1) & 3.1(1) & 1.2(1)\\
\tableline
\tableline 
\end{tabular} ~\vspace*{\baselineskip}~\\ \end{center}
  \tablecomments{$R_{\rm in}$ values in $R_{\rm g}$.  In the VHS, the Laor
   relativistic disk line emission model can be fit to the
   Fe~K$_{\alpha}$ profile with $R_{\rm in} = 1.2~R_{\rm g}$ with statistical
   significance comparable to that of the Gaussian model.\\ 
   The relation described by DiMatteo \& Psaltis (1999) is shown in
   Equation 1 within the text.  Given $\nu_{QPO}$ and assuming a BH
   mass, an inner disk radius may be derived assuming Keplerian orbits
   up to the marginally stable orbit.\\
   (a) QPO frequencies reported by Revnivtsev et al.\ 1999.\\ 
   (b) Calculated via Equation 1 for $M_{BH}=5.0~\Msun$.\\
   (c) Calculated via Equation 1 for $M_{BH}=10.0~\Msun$.\\
   (d) The multicolor blackbody disk radius.\\}
\vspace{-1.0\baselineskip}
\end{table}

\nopagebreak

\begin{table}[t]
\caption{LS observations fit with the Diskline model} \scriptsize
\begin{center} \begin{tabular}{lllllll} 
Obs. & $E_{cent}$ & $R_{\rm in}$ & EW & Flux(3--25 keV) & $\chi^{2}$/dof & $\chi^{2}$/dof \\ 
~ & (keV) & ($R_{\rm g}$) & (eV) & $10^{-11}~cgs$ & (Gaussian) & (diskline)\\ 
\tableline 
\tableline 
16 & $6.68(4)$ & $63^{+933}_{-31}$ & $200(12)$ & $2.1(1)$ & 0.75 & 0.55 \\
17 & $6.72(3)$ & $25^{+26}_{-19}$ & $411(16)$ & $2.5(1)$ & 0.62 & 0.77 \\
18 & $6.69(3)$ & $6^{+15}$ & $533(19)$ & $2.6(1)$ & 1.02 & 0.81 \\
19 & $6.70(4)$ & $102^{+140}_{-63}$ & $525(24)$ &$2.6(1)$ & 0.85 & 0.91 \\
20 & $6.81(4)$ & $22^{+33}_{-16}$ & $612(29)$ & $1.8(1)$ & 0.78 & 0.89 \\
21 & $6.78(3)$ & $25^{+29}_{-19}$ & $682(27)$ & $1.8(1)$ & 0.72 & 0.80 \\
22 & $6.65(4)$ & $14^{+12}_{-8}$ & $915(41)$ & $2.9(1)$ & 0.77 & 0.96 \\
23 & $6.79(3)$ & $35^{+34}_{-15}$ & $916(33)$ & $2.0(1)$ & 0.40 & 0.80 \\
\tableline \tableline
\end{tabular} ~\vspace*{\baselineskip}~\\ \end{center}
  
  \tablecomments{We fix $R_{out}$ at $10^{4}~R_{\rm g}$, the emissivity
  profile to $r^{-2}$, and the disk inclination at 45 degrees.  With
  these parameters fixed, and using the pexriv parameters obtained via
  fitting with a Gaussian line model, we measure line centroid energy
  ($E_{cent}$), inner radius ($R_{\rm in}$), equivalent width (EW), the
  3.5-20.0 keV line flux, and the reduced $\chi^{2}$ values for fits
  with a Gaussian, and with diskline.  Errors are 90\% confidence
  limits. Fits have 36 dof without the diskline model, 33 dof with the
  model.  All fits with the diskline model are significant at greater
  than 99.9\% confidence.  Although the relative $\chi^{2}$ values do
  not demand the diskline over the Gaussian model, the parameters of
  the diskline model are very well-constrained.}
\vspace{-1.0\baselineskip}
\end{table}

\clearpage

\begin{figure}
\figurenum{2}
\label{fig:VHS fitting}
\centerline{~\psfig{file=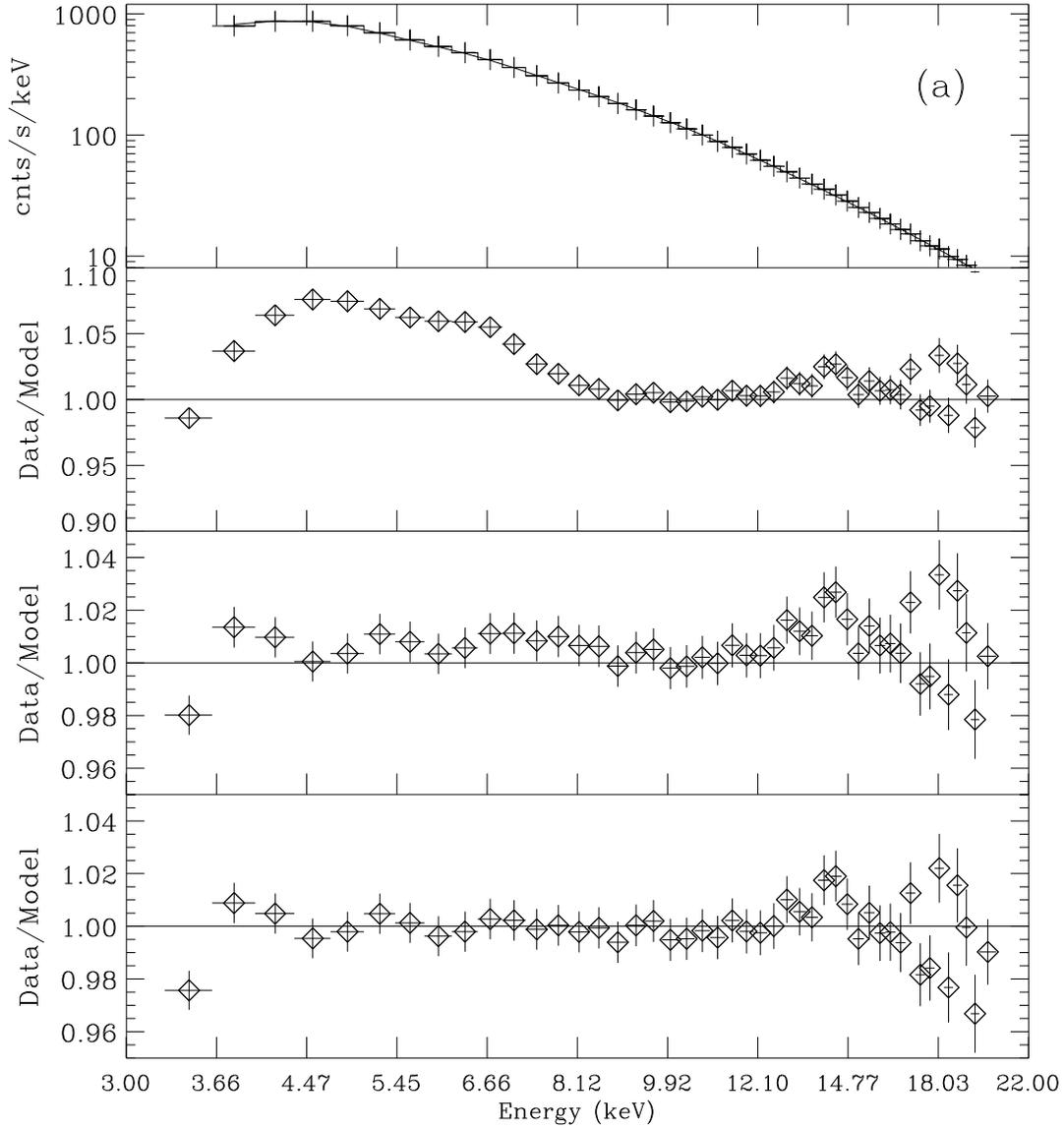,width=6.0in,height=7.0in}~}
\caption{Typical spectra and data/model profiles from the VHS and LS.
\scriptsize (a) Observation 3 (VHS).  The top plot shows the spectrum
for Observation 3, fit with a blackbody disk + 2 Gaussians + Pexriv
model.  The plots below are data/model ratio plots.  Second from top,
dbb + plaw; notice the double line peaks and the reflection above 10
keV.  The third plot is the ratio for a model consisting of dbb + 2
Gaussians + plaw; the lines are fit well but the reflection is not.
Finally, at bottom, the ratio for a model consisting of dbb + two gs +
pexriv; this model fits the lines and continuum very well. (b)
Observation 19 (LS).  The top plot shows the spectrum for Observation
19, fit with a 2 Gaussians plus Pexriv model.  The plots below are
data/model ratio plots.  Second from top, p-law; notice the very
strong, broadened line and reflection.  The third plot is the ratio
for a model consisting of a Gaussian + plaw; the lines is fit well,
but a higher energy line remains.  Fourth, the ratio for a model with
2 g's + plaw; this is a good fit.  At bottom, the ratio for a model
consisting of two g's + reflection via pexriv.}
\end{figure}

\clearpage

\begin{figure}
\figurenum{2}
\label{fig:LS fitting}
\centerline{~\psfig{file=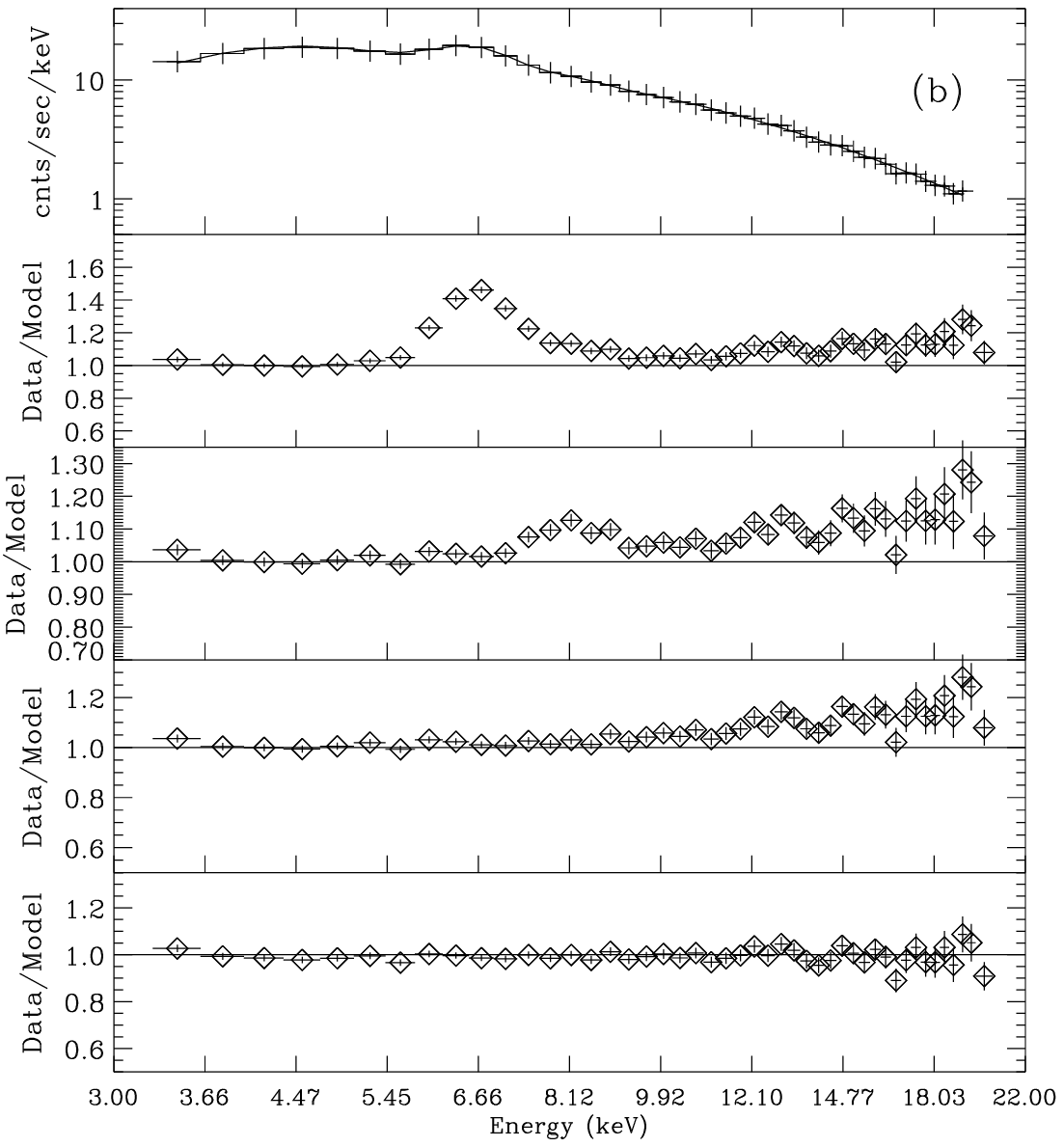,width=6.0in,height=7.0in}~}
\end{figure}

\clearpage

\begin{figure}
\figurenum{3}
\label{fig:VHS and HS Ratios }
\centerline{~\psfig{file=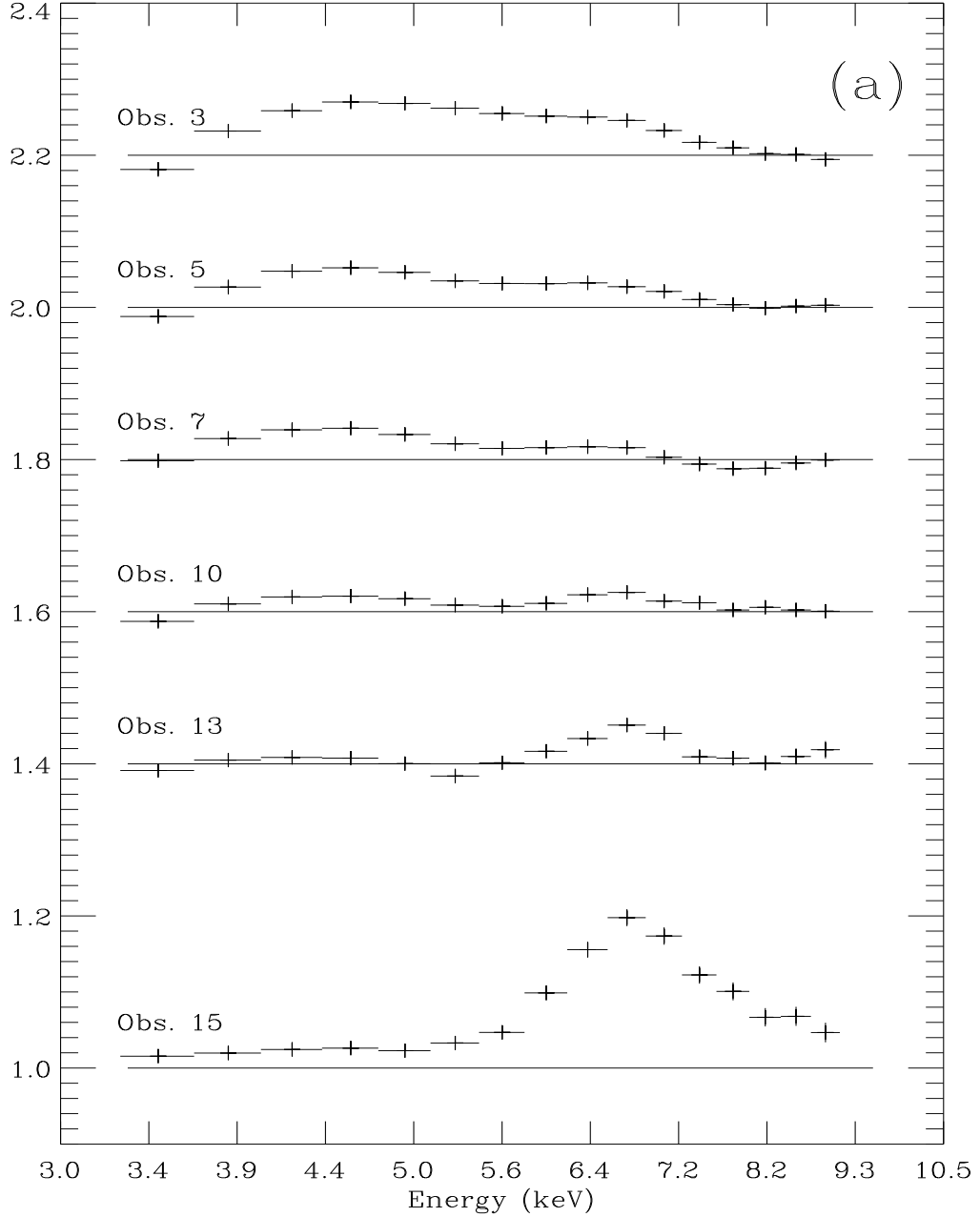,width=6.0in,height=7.0in}~}
\caption{Data/model ratios across the outburst.  \scriptsize(a) VHS
and HS: Models include only the basic multicolor blackbody disk plus
power-law components.  Observations are offset with respect to each
other for clarity; however, the absolute scale is teh same for all.
The emission excess in observation 3 (at top) is double-peaked, but by
the middle of the HS it is clear the profile is becoming
single-peaked.  Finally, in observation 15 (at bottom) the line
feature is very strong and broad.  (b) LS: The model is the basic
power-law only.  It is clear that the line is very prominent relative
to the continuum in the LS.  Some variation in line centroid energies
and widths can be discenered.}
\end{figure}
\nopagebreak

\clearpage

\begin{figure}
\figurenum{3}
\label{fig:LS ratios}
\centerline{~\psfig{file=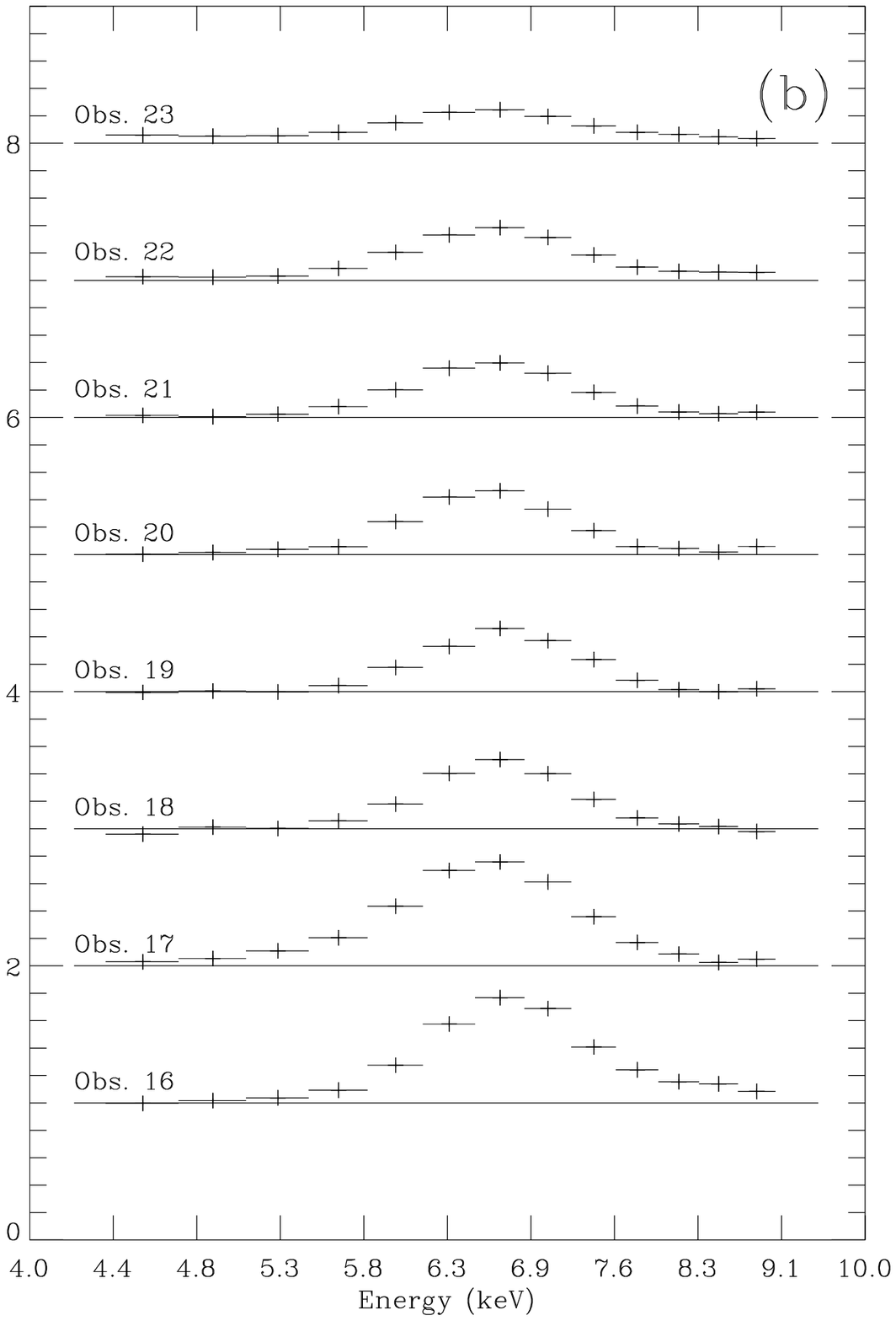,width=6.0in,height=7.0in}~}
\end{figure}
\nopagebreak

\clearpage

\begin{figure}
\figurenum{4}
\label{fig:VHS and HS values}
\centerline{~\psfig{file=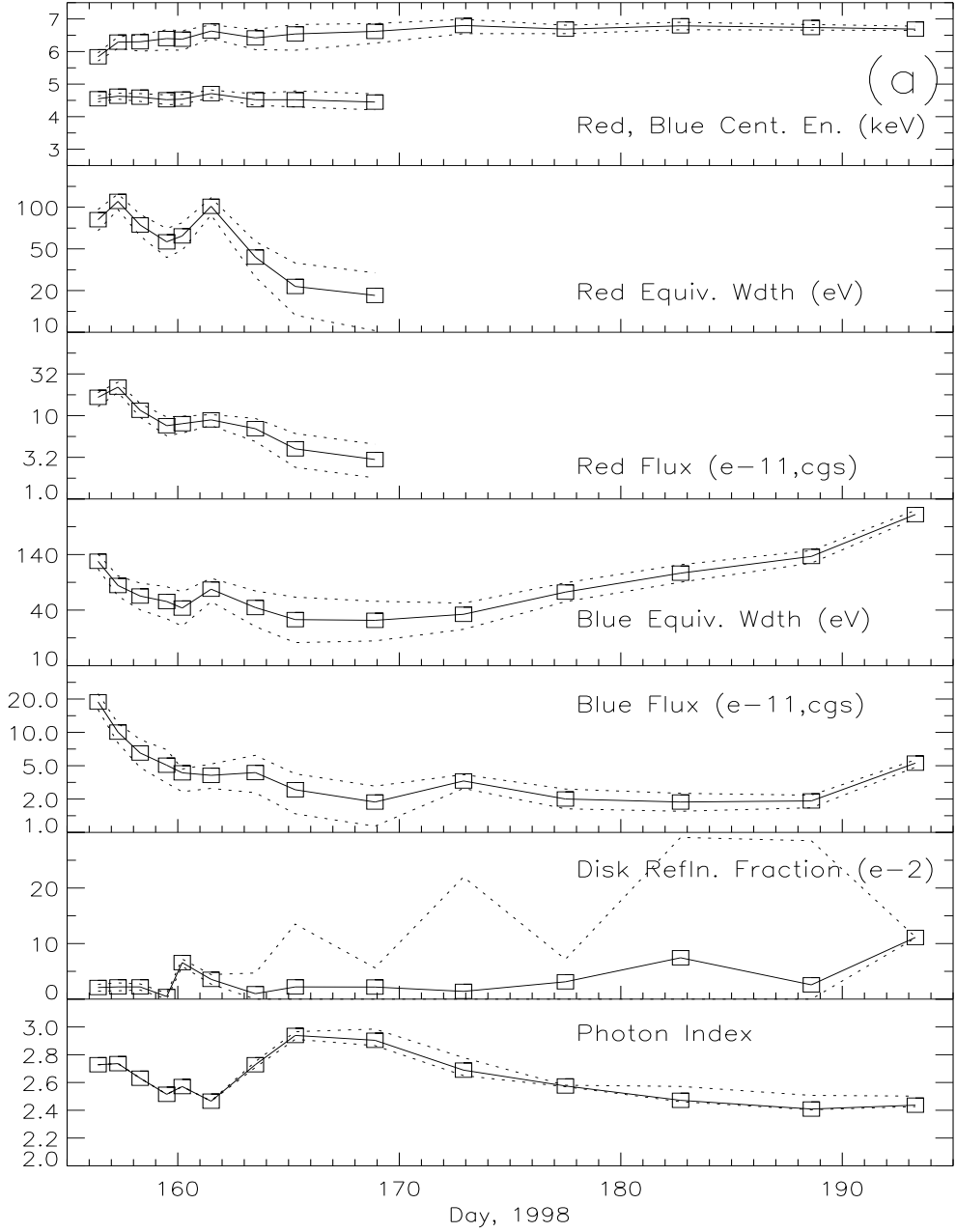,width=6.0in,height=7.0in}~}
\caption{Line and Reflection Parameters. \scriptsize Measured values
are connected by a continuous line, and 90\% confidence limits by
dotted lines.  (a) VHS and HS.  Top to bottom: red and blue wing
centroid energy (top); red wing equivalent width; red wing flux; blue
line equivalent width; blue wing flux; disk reflection fraction; and
power-law photon index (bottom); vs Day of Outburst (days from 3.8
June 1998).  The disk ionization parameter is high throughout the VHS and
HS; $\xi~\geq~2.0\times10^{3}$. (b) LS.  Top to bottom: Fe~K$_{\alpha}$
centroid energy (top); Fe~K$_{\alpha}$ equivalent width;
Fe~K$_{\alpha}$ flux; disk reflection fraction; and
power-law photon index (bottom); vs Day of Outburst.  The disk
ionization parameter is low in the LS; $0~\leq~\xi~\leq~100$. }
\end{figure}
\nopagebreak

\clearpage

\begin{figure}
\figurenum{4}
\label{fig:LS values}
\centerline{~\psfig{file=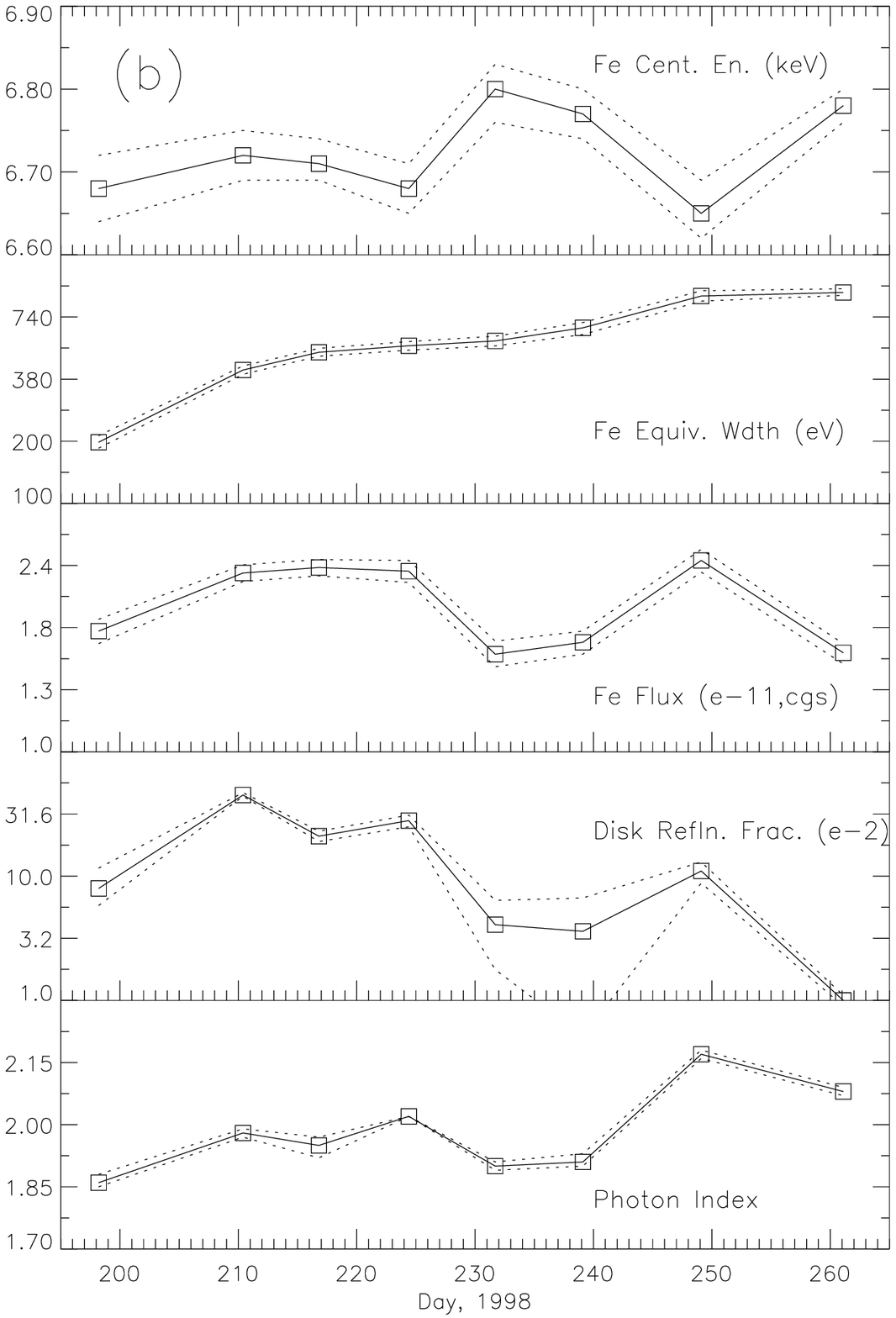,width=6.0in,height=7.0in}~}
\end{figure}
\nopagebreak

\end{document}